\documentclass[aps,prb,showpacs,groupedaddress,twocolumn]{revtex4}
\usepackage{bm,amsmath}
\usepackage[dvips]{graphicx,color}
\begin{document}

\title{Mesoscopic phonon transmission through a nanowire-bulk contact}

\author{Chun-Min Chang and Michael R. Geller}
\affiliation{Department of Physics and Astronomy, University of Georgia, Athens, Georgia 30602-2451}
\date{September 23, 2004}

\begin{abstract}
We calculate the frequency-dependent mesoscopic acoustic phonon transmission probability through the abrupt junction between a semi-infinite, one-dimensional cylindrical quantum wire and a three-dimensional bulk insulator, using a perturbative technique that is valid at low frequency. The system is described using elasticity theory, and traction-free boundary conditions are applied to all free surfaces. In the low-frequency limit the transmission probability vanishes as $\omega^2 \! ,$ the transport being dominated by the longitudinal channel, which produces a monopole source of elastic radiation at the surface of the bulk solid. The thermal conductance between an equilibrated wire nonadiabatically coupled to a bulk insulator should therefore vanish with temperature as $T^3 \! .$
\end{abstract}


\maketitle
\clearpage

\section{INTRODUCTION}

Electronic transport in a variety mesoscopic systems has been successfully described by the theory of Landauer and B\"{u}ttiker.\cite{Landauer,Buttiker,van Wees etal,Wharam etal,Beenakker review,Buttiker review,Datta book}  In this scattering approach, each fully propagating channel in a wire contributes $2 \pi e^2/\hbar$ to the electrical conductance. Recently there have been experimental efforts to study phonons in the mesoscopic regime,\cite{Roukes phonon counting} and beautiful experiments by Schwab {\it et al.}\cite{Schwab etal quantization} and by Yung {\it et al.}\cite{Yung etal quantization} have observed the low-temperature universal thermal conductance quantum of $\pi k_{B}^{2}T/6\hbar$ per vibrational channel. In these experiments, the nanowires were adiabatically (on the scale of the thermal wavelength) connected to thermalized phonon reservoirs,\cite{adiabatic footnote} and the observation of a thermal conductance varying linearly with temperature implies that very little phonon reflection occurred at the nanowire-bulk junctions.

In the presence of an abrupt, nonadiabiatic coupling between the nanowire and bulk reservoirs, phonons will scatter at the junctions and suppress the thermal conductance.\cite{Schwab etal quantization,Rego and Kirczenow,Cross and Lifshitz,Patton and Geller} Cross and Lifshitz\cite{Cross and Lifshitz} have calculated the frequency-dependent acoustic phonon transmission probability ${\sf T}$ between a semi-infinite quantum wire of rectangular cross-section and a thin plate with the same thickness as the wire, and find that ${\sf T} \propto \omega^{1/2}$ in the low-frequency limit. In Ref.~[\onlinecite{Patton and Geller}] a short nanowire, modeled as a harmonic spring, abruptly connected at both ends to three-dimensional bulk insulators was found to have ${\sf T} \propto \omega^2.$ These result suggest that such nanowires will eventually become thermal {\it insulators} at low temperatures, and the universal thermal conductance quantum will not be observed. However, such a cross-over to insulating behavior has not yet been observed experimentally.

Understanding the scattering caused by nonadiabatic nanowire-bulk contacts will be essential for pushing phonon physics into the mesoscopic regime, as well as for the design and operation of thermal nanosensors such as calorimeters and bolometers. In this paper, we extend the previous work by calculating the mesoscopic acoustic phonon transmission probability through the abrupt junction between a semi-infinite, one-dimensional cylindrical quantum wire and a three-dimensional bulk insulator, using the low-frequency perturbative technique employed by Cross and Lifshitz.\cite{Cross and Lifshitz} The nanowire and bulk insulators are described using isotropic elasticity theory, and traction-free boundary conditions are applied to all free surfaces. In the low-frequency limit the transmission probability is found to vanish as $\omega^2 \! ,$ with the low-frequency transport being dominated by the longitudinal channel. 

In the Sec.~\ref{wire modes section} we review the calculation of the long-wavelength vibrational modes of an infinitely long cylindrical elastic rod. The long wavelength limit of interest here is defined as $k b \ll 1$, where $b$ is the radius of the cylinder and $k$ is the wavenumber. In Sec.~\ref{boundary conditions section} we show that in the long-wavelength limit, the bulk solid produces a hard-wall boundary condition on the nanowire. In Sec.~\ref{response function section} we calculate the displacement field in the three-dimensional bulk solid given an applied traction to its surface, using what is essentially an elastic Green's function method.\cite{Cross and Lifshitz,Miller and Pursey} The frequency-dependent transmission probabilities for each of the four gapless modes are calculated in Sec.~\ref{transmission probability section}, and our conclusions are given in Sec.~\ref{conclusion section}.

\section{VIBRATIONAL MODES OF CYLINDRICAL WIRE\label{wire modes section}}

In this section, we will briefly review the elastic waves for an infinitely long cylindrical waveguide. In the long wavelength limit there are four branches, which include one torsional branch, one longitudinal branch, and two flexural branches.\cite{Graff book} We assign a numerical subscript to represent each branch mode, with ``1" denoting the torsional branch, ``2" denoting the longitudinal branch, and ``3" and ``4" denoting the flexural branches. These four branches are orthogonal. Also, cylindrical coordinates are used below.

We assume an isotropic elastic continuum with transverse and the longitudinal sound speeds 
\begin{equation}
c_{t}=\sqrt{\frac{\mu }{\rho }} \ \ \ \ {\rm and} \ \ \ \ c_{l}=\sqrt{\frac{\lambda +2\mu }{\rho }},
\end{equation}
where $\rho$ is the mass density, and $\lambda $ and $\mu$ are the Lam\'{e} constants.

\subsection{Branch 1: torsional}

The displacement field is given by
\begin{equation}
\mathbf{u}_{1}(\mathbf{r},t) =r\text{ }e^{i\left( kz-\omega_{1}t\right) }\mathbf{e}_{\theta },
\end{equation}
with dispersion relation
\begin{equation}
\omega _{1}=c_{t} k.
\end{equation}
The stress tensor elements acting on a cross-section of the rod are
\begin{eqnarray}
\sigma _{1rz} &=&\mu \left( \frac{\partial u_{1r}}{\partial z}+\frac{\partial u_{1z}}{\partial r}\right) = 0,  \notag \\
\sigma _{1\theta z} &=&\mu \left( \frac{\partial u_{1\theta }}{\partial z}+\frac{1}{r}\frac{\partial u_{1z}}{\partial \theta }\right) =i\mu kre^{i\left( kz-\omega _{1}t\right) }, \notag \\
\sigma _{1zz} &=&\left[ \lambda \left( \nabla \cdot \bf{u}_{1}\right) +2\mu \frac{\partial u_{1z}}{\partial z}\right] =0.
\label{torsional stress tensors}
\end{eqnarray}

\subsection{Branch 2: longitudinal }

The displacement field is 
\begin{equation}
\mathbf{u}_{2}(\mathbf{r},t) =\left[ f_{r}(r)\mathbf{e}_{r}+f_{z}(r)\mathbf{e}_{z}\right] e^{i\left( kz-\omega _{2}t\right) },
\end{equation}
where
\begin{eqnarray}
f_{r}(r) &=&-A_{2}\alpha J_{1}(\alpha r)+B_{2}ikJ_{1}(\beta r),  \notag \\
f_{z}(r) &=&A_{2}ikJ_{0}(\alpha r)-B_{2}\beta J_{0}(\beta r).
\end{eqnarray}
$J_{n}(r)$ is the Bessel function of $n$th order. $\alpha$ and $\beta$ are constants determined by the boundary conditions. Furthermore,
\begin{equation}
\alpha =\sqrt{\frac{\omega _{2}{}^{2}}{c_{l}{}^{2}}-k^{2}} \ \ \ {\rm and} \ \ \ \beta =\sqrt{\frac{\omega _{2}{}^{2}}{c_{t}{}^{2}}-k^{2}},
\end{equation}
and
\begin{equation}
\frac{A_{2}}{B_{2}}=\frac{\beta ^{2}-k^{2}}{2i\alpha k}\frac{J_{1}(\beta b)}{J_{1}(\alpha b)}.
\end{equation}
The relevant stresses are
\begin{eqnarray}
\sigma _{2rz} &=&\mu \left[ ikf_{r}(r)+\frac{df_{z}(r)}{dr}\right] e^{i\left( kz-\omega _{2}t\right) } , \notag \\ 
\sigma _{2\theta z} &=&0 , \notag \\
\sigma _{2zz} &=& \bigg[\lambda \left( \frac{df_{r}(r)}{dr}+\frac{f_{r}(r)}{r}\right)  \notag \\
&&+ik\left( \lambda +2\mu \right) f_{z}(r) \bigg]e^{i\left( kz-\omega _{2}t\right)}.
\end{eqnarray}

In the long wavelength $kb \ll 1$ limit, $\omega _{2}=c_{0}k,$
\begin{equation}
\alpha =ik\sqrt{1-\frac{c_{0}{}^{2}}{c_{l}{}^{2}}}, \ \ \ {\rm and} \ \ \ \beta =k \sqrt{\frac{c_{0}{}^{2}}{c_{t}{}^{2}}-1},
\end{equation}
where $c_{0}$ is related to Young's modulus of elasticity $E$ by
\begin{equation}
c_{0}=\sqrt{\frac{E}{\rho }}=c_{t}\sqrt{\frac{3c_{l}{}^{2}-4c_{t}{}^{2}}{c_{l}{}^{2}-c_{t}{}^{2}}}.
\end{equation}
To leading order
\begin{equation}
\mathbf{u}_{2}\left( \mathbf{r},t\right) = \mathbf{e}_{z} \, e^{i\left( kz-\omega _{2}t\right) },
\end{equation}
and the stresses are
\begin{eqnarray}
\sigma _{2rz} &=& 0,  \notag \\
\sigma _{2\theta z} &=& 0,  \notag \\
\sigma _{2zz} &=&i\mu \frac{c_{0}{}^{2}}{c_{t}{}^{2}}k\text{ }e^{i\left(kz-\omega _{2}t\right) }  \notag \\
&=&i\mu \left( \frac{3-4p^{2}}{1-p^{2}}\right) k\text{ }e^{i\left( kz-\omega_{2}t\right) },
\end{eqnarray}
where $p \equiv c_{t}/c_{l}$ is the ratio of the transverse to longitudinal sound speed.

\subsection{Branch 3: $x$-polarized flexural}

The displacement field is
\begin{eqnarray}
\mathbf{u}_{3}(\mathbf{r},t) &=& \bigg[ g_{r}(r)\cos \theta \, \mathbf{e}_{r}+g_{\theta }\left( r\right) \sin \theta \, \mathbf{e}_{\theta} \notag  \\
&+& g_{z}(r)\cos \theta \, \mathbf{e}_{z} \bigg] e^{i\left( kz-\omega_{3}t \right)},
\end{eqnarray}
where
\begin{eqnarray}
g_{r}(r) &=&\frac{dJ_{1}(\alpha r)}{dr}+B_{3}\frac{dJ_{1}(\beta r)}{dr}+C_{3} \frac{J_{1}(\beta r)}{r},  \notag \\
g_{\theta }\left( r\right) &=&-\frac{J_{1}(\alpha r)}{r}-B_{3}\frac{J_{1}(\beta r)}{r}-C_{3}\frac{dJ_{1}(\beta r)}{dr},  \notag \\
g_{z}(r) &=&ikJ_{1}(\alpha r)-iB_{3}\frac{\beta ^{2}}{k}J_{1}(\beta r).
\end{eqnarray}
$B_{3}$ and $C_{3}$ are constants. The stresses are given by
\begin{eqnarray}
\sigma _{3rz} &=&\mu \left[ ikg_{r}(r)+\frac{dg_{z}(r)}{dr}\right] \cos \theta \, e^{i\left( kz-\omega _{3}t\right) } , \notag \\
\sigma _{3\theta z} &=&\mu \left[ ikg_{\theta }(r)-\frac{g_{z}(r)}{r}\right] \sin \theta \, e^{i\left( kz-\omega _{3}t\right) } , \notag \\
\sigma _{3zz} &=& \bigg[ \lambda \left( \frac{dg_{r}(r)}{dr}+\frac{g_{r}(r)+g_{\theta }(r)}{r}\right)   \notag \\
&+& i \left( \lambda +2\mu \right) k \, g_{z}(r) \bigg] \cos \theta \, e^{i\left( kz-\omega _{3}t\right) }.
\end{eqnarray}

In the $kb<<1$ limit,
\begin{equation}
\omega _{3}=\frac{c_{0}}{2}bk^{2},
\end{equation}
\begin{equation}
\alpha =ik\sqrt{1-\left( \frac{c_{0}bk}{2c_{l}}\right) ^{2}}, 
\end{equation}
and
\begin{equation}
\beta =ik\sqrt{1-\left( \frac{c_{0}bk}{2c_{t}}\right) ^{2}}.
\end{equation}
To leading order,
\begin{eqnarray}
\mathbf{u}_{3}\left( \mathbf{r},t\right) &=&\left( \cos \theta \mathbf{e}_{r}-\sin \theta \mathbf{e}_{\theta }+ikx\mathbf{e}_{z} \right) e^{i\left( kz-\omega _{3}t\right) }  \notag \\
&=&\left( \mathbf{e}_{x}-ikx\mathbf{e}_{z} \right) e^{i\left( kz-\omega _{3}t\right) }  ,
\label{flexural displacement}
\end{eqnarray}
and the stresses are
\begin{eqnarray}
\sigma _{3xz} &=&\frac{i\mu }{4}\left[ \left( 1+\frac{c_{0}{}^{2}}{c_{t}{}^{2}}\right) \left( b^{2}-x^{2}\right) \right.  \notag \\
&&\left. -\left( 3-\frac{c_{0}{}^{2}}{c_{t}{}^{2}}\right) y^{2}\right]
k^{3}e^{i\left( kz-\omega _{3}t\right) } , \notag \\
\sigma _{3yz} &=&\frac{i\mu }{2}\left( 1-\frac{c_{0}{}^{2}}{c_{t}{}^{2}}\right) k^{3}xye^{i\left( kz-\omega _{3}t\right) }  , \notag \\
\sigma _{3zz} &=&\mu \frac{c_{0}{}^{2}}{c_{t}{}^{2}}k^{2}x\text{ }e^{i\left(kz-\omega _{3}t\right) }.
\label{flexural stress tensors}
\end{eqnarray}
The rod bends in the $xz$ plane.

\subsection{Branch 4: $y$-polarized flexural}

An independent flexural mode can be found by letting the rod bend in the $y$ direction. In the long-wavelength limit,
\begin{equation}
\mathbf{u}_{4}\left( \mathbf{r},t\right) =\left( \mathbf{e}_{y}-iky\mathbf{e}_{z} \right) e^{i\left( kz-\omega _{3}t\right) },
\end{equation}
and the stresses on the surface normal to $z$ are
\begin{eqnarray}
\sigma _{4yz} &=&\frac{i\mu }{4}\left[ \left( 1+\frac{c_{0}{}^{2}}{c_{t}{}^{2}}\right) \left( b^{2}-y^{2}\right) \right.  \notag \\
&&\left. -\left( \frac{c_{0}{}^{2}}{c_{t}{}^{2}}-3\right) x^{2}\right] k^{3}e^{i\left( kz-\omega _{3}t\right) } , \notag \\
\sigma _{4xz} &=&-\frac{i\mu }{2}\left( \frac{c_{0}{}^{2}}{c_{t}{}^{2}}-1\right) k^{3}xye^{i\left( kz-\omega _{3}t\right) } , \notag \\
\sigma _{4zz} &=&\mu \frac{c_{0}{}^{2}}{c_{t}{}^{2}}k^{2}y\text{ }e^{i\left(kz-\omega _{3}t\right) }.
\end{eqnarray}

\section{BOUNDARY CONDITIONS AT THE NANOWIRE-BULK INTERFACE\label{boundary conditions section}}

The essence of the perturbative method we use is as follows: In an abrupt junction geometry, the bulk solid presents a stiff boundary to the nanowire, so to zeroth order one calculates the vibrational modes of the wire assuming a zero-displacement boundary condition at the contact. The stress distributions associated with these vibrational modes is then calculated in the junction region. These zeroth-order vibrational modes, however, do not carry elastic energy because of the hard-wall boundary condition associated with the infinitely stiff bulk solid. Now one relaxes the hard-wall boundary condition, replacing it with the condition that elastic waves in the bulk are purely radiative, having outward moving components but no inward ones. The elastic energy radiated by the nanowire into the semi-infinite bulk solid is then computed using the actual elastic parameters of the bulk, and the ratio of incident to transmitted energy determines the transmission probability.

Thus, as explained, we will calculate the elastic stress on the surface of the three-dimensional bulk insulator, produced by the vibrating nanowire, by assuming that the bulk imposes a zero-displacement boundary condition on the nanowire. This is physically reasonable, and can be further justified by considering the bulk to be a thick wire with a radius $B$ much larger than the nanowire radius $b.$\cite{Cross and Lifshitz} Assuming a sound wavelength larger than both $b$ and $B$, the conservation of linear and angular momentum lead the zero-displacement boundary condition in the limit $B \gg b$.

We consider a semi-infinite cylindrical elastic nanowire of radius $b$, lying along the $z$ axis and connected at $z=0$ to a thicker cylinder of radius $B$. An incident elastic wave $\mathbf{u}_{i}=\mathbf{u}_{i}\left(r,\theta \right) e^{i\left( kz-\omega _{3}t\right) }$ propagates from the nanowire to thick cylinder. $k$ is smaller than both $b$ and $B$ so both cylinders are still one-dimensional, and $i=1,2,3,4$ denotes the branch. The scattering causes a reflected wave for $z<0$ and a transmitted wave for $z>0$. We can write the displacement field as
\begin{equation}
\begin{array}{cc}
\mathbf{u}_{i}\left( r,\theta \right) e^{i\left( kz-\omega t\right) }+R_{ij}
\mathbf{u}_{j}^{\ast }\left( r,\theta \right) e^{-i\left( kz+\omega t\right)} & \text{for} \ z<0 \\ &  \\ 
T_{ij}\mathbf{u}_{j}\left( r,\theta \right) e^{i\left( kz-\omega t\right) } & \text{for} \ z>0.
\end{array}   
\label{displacement of the system}
\end{equation}
Here $R_{ij}$ and $T_{ij}$ are the reflection and transmission coefficients, which are matrices in the channel indices. 

The continuity of the displacement field, combined with the orthogonality of the vibrational eigenmodes, leads to
\begin{equation}
\delta _{ij}+R_{ij} = T_{ij}.  
\label{contiuum of displacement vector}
\end{equation}
In the Appendix we show that in the $B \gg b$ limit, conservation of linear and angular momentum leads to 
\begin{equation}
R_{ij} \rightarrow -\delta_{ij} \ \ \ {\rm and} \ \ \ T_{ij} \rightarrow 0,
\label{hard-wall boundary conditions}
\end{equation}
which means that elastic waves are flipped upon reflection, and no interbranch scattering occurs. This result is analogous to that obtained by Cross and Lifshitz in their thin-plate geometry.\cite{Cross and Lifshitz}

Linear combinations of the vibrational eigenfunctions of Sec.~\ref{wire modes section} can be used to satisfy the boundary conditions of Eq.~(\ref{hard-wall boundary conditions}). These displacement fields produce the following stress distributions at the $z=0$ interface: For the torsional mode we obtain
\begin{equation}
\sigma _{\theta z}=\left\{ \begin{array}{cc}
2i\mu kr^{2}e^{-i\omega t} & \text{for} \ r<b, \\ 
&  \\ 0 & \text{for} \ r>b.
\end{array} 
\right.  \label{torsional BC}
\end{equation}
For the longitudinal mode,
\begin{equation}
\sigma _{zz}=\left\{ 
\begin{array}{cc}
2i\mu \frac{c_{0}{}^{2}}{c_{t}{}^{2}}ke^{-i\omega t} & \text{for} \ r<b,
\\ &  \\ 0 & \text{for} \ r>b.
\end{array}
\right.  
\label{longitudinal BC}
\end{equation}
And for the $x$-polarized flexural mode, we find
\begin{equation}
\begin{array}{c}
\sigma _{xz}=\sigma _{yz}=\sigma _{zz}=0
\end{array}
\ \ \ \ \ \ \ \ \ \ \ \text{for} \ r>b,  
\label{flexural BC0}
\end{equation}
\begin{equation}
\begin{array}{ll}
\sigma _{xz}= & \frac{i\mu }{2}\left[ \left( 1+\frac{c_{0}{}^{2}}{c_{t}{}^{2}
}\right) \left( b^{2}-x^{2}\right) \right. \\ 
&  \\ 
& -\left. \left( 3-\frac{c_{0}{}^{2}}{c_{t}{}^{2}}\right) y^{2}\right]
k^{3}e^{-i\omega _{3}t} \\  &  \\ 
\sigma _{yz}= & i\mu \left( 1-\frac{c_{0}{}^{2}}{c_{t}{}^{2}}\right)
k^{3}xye^{-i\omega _{3}t} \\  &  \\ 
\sigma _{zz}= & 0
\end{array}
\text{for} \ r<b.  \label{flexural BC}
\end{equation}
The stress distribution from the $y$-polarized flexural mode has the same form as (\ref{flexural BC0}) and (\ref{flexural BC}) after exchanging $x\leftrightarrow y.$

\section{3D ELASTIC RESPONSE FUNCTION\label{response function section}}

Next we calculate the displacement field in the three-dimensional solid given the applied stress of Sec.~\ref{boundary conditions section} to its surface at $z=0$. For $r < b$, this is the stress distribution produced by the nanowire, and for $r>b$ it is the stress imposed by the traction-free boundary condition. The method we use here, which is essentially a Green's function method, is well known in elasticity theory.\cite{Cross and Lifshitz,Miller and Pursey}

To find the displacement field in the bulk solid given the boundary conditions described above, a scalar potential $\phi $ and a vector potential $\mathbf{H}$ are introduced according to
\begin{equation}
\mathbf{u}=\mathbf{\nabla }\phi +\mathbf{\nabla }\times \mathbf{H.}
\label{displacement vector}
\end{equation}
The wave equations for the potential fields are
\begin{equation}
\left( \frac{\partial ^{2}}{\partial t^{2}}-c_{l}{}^{2}\mathbf{\nabla }^{2}\right) \phi =0,\text{ \ \ }\left( \frac{\partial ^{2}}{\partial t^{2}}
-c_{t}{}^{2}\mathbf{\nabla }^{2}\right) \mathbf{H}=0.  \label{wave equations}
\end{equation}
They can be written as
\begin{equation}
\phi \left( x,y,z\right) = \frac{1}{2\pi }\int_{-\infty }^{\infty }dk_{1}dk_{2}f\left( k_{1},k_{2}\right) e^{-i\left( k_{1}x+k_{2}y\right) }e^{ik_{l3}z} 
\end{equation}
and
\begin{equation}
\mathbf{H}\left( x,y,z\right)  = \frac{1}{2\pi }\int_{-\infty }^{\infty}dk_{1}dk_{2}\mathbf{h}\left( k_{1},k_{2}\right) e^{-i\left(k_{1}x+k_{2}y\right) }e^{ik_{t3}z},
\end{equation}
where
\begin{equation}
k_{l3}=\sqrt{\frac{\omega ^{2}}{c_{l}{}^{2}}-k_{1}{}^{2}-k_{2}{}^{2}} 
\end{equation}
and
\begin{equation}
 k_{t3}=\sqrt{\frac{\omega ^{2}}{c_{t}{}^{2}} -k_{1}{}^{2}-k_{2}{}^{2}}.
\end{equation}
Here $f$  and $\mathbf{h}$ are the Fourier transforms of the potential fields $\phi$ and  $\mathbf{H}$ at $z=0$. 

Now, we can use Eq.~(\ref{displacement vector}) and choose the transverse ``gauge'' $\mathbf{\nabla }\cdot \mathbf{H}=0$ to express the components of the displacement vector by the inverse Fourier transform $\mathcal{F}^{-1}$,
\begin{eqnarray}
u_{x}\left( x,y,z\right)  &=&-i\mathcal{F}^{-1}\left[ k_{1}f\left(
k_{1},k_{2}\right) e^{ik_{l3}z}+g_{x}\left( k_{1},k_{2}\right) e^{ik_{t3}z}%
\right]   \notag \\
u_{y}\left( x,y,z\right)  &=&-i\mathcal{F}^{-1}\left[ k_{2}f\left(
k_{1},k_{2}\right) e^{ik_{l3}z}+g_{y}\left( k_{1},k_{2}\right) e^{ik_{t3}z}%
\right]   \notag \\
u_{z}\left( x,y,z\right)  &=& i\mathcal{F}^{-1}\bigg[ k_{l3}f\left(
k_{1},k_{2}\right) e^{ik_{l3}z}  \notag \\
&-& \frac{k_{1}g_{x}\left( k_{1},k_{2}\right) +k_{2}g_{y}\left(
k_{1},k_{2}\right) }{k_{t3}}e^{ik_{t3}z}\bigg] ,
\label{components of displacement}
\end{eqnarray}
where
\begin{eqnarray}
g_{x}\left( k_{1},k_{2}\right)  &=&k_{2}h_{z}\left( k_{1},k_{2}\right)
+k_{t3}h_{y}\left( k_{1},k_{2}\right)  \\
g_{y}\left( k_{1},k_{2}\right)  &=&-k_{1}h_{z}\left( k_{1},k_{2}\right)
-k_{t3}h_{x}\left( k_{1},k_{2}\right) .
\end{eqnarray}

The stress at the boundary $z=0$ can also be expressed in terms of the inverse Fourier transforms, as
\begin{eqnarray}
\sigma _{xz} &=&\mu \left( \frac{\partial u_{x}}{\partial z}+\frac{\partial
u_{z}}{\partial x}\right) _{z=0}  \notag \\
&=&\mu \mathcal{F}^{-1} \bigg[ 2k_{1}k_{l3}f\left( k_{1},k_{2}\right) \notag \\
&+& \frac{\left( k_{t3}{}^{2}-k_{1}{}^{2}\right) g_{x}\left(k_{1},k_{2}\right) -k_{1}k_{2}g_{y}\left( k_{1},k_{2}\right) }{k_{t3}} \bigg], \notag \\
\sigma _{yz} &=&\mu \left( \frac{\partial u_{y}}{\partial z}+\frac{\partial u_{z}}{\partial y}\right) _{z=0}  \notag \\
&=&\mu \mathcal{F}^{-1} \bigg[ 2k_{2}k_{l3}f\left( k_{1},k_{2}\right)  \notag \\
&+& \frac{\left( k_{t3}{}^{2}-k_{2}{}^{2}\right) g_{y}\left( k_{1},k_{2}\right) -k_{1}k_{2}g_{x}\left( k_{1},k_{2}\right) }{k_{t3}} \bigg], \notag \\
\sigma _{zz} &=&\left[ \lambda \left( \mathbf{\nabla }\cdot \vec{u}\right) +2\mu \frac{\partial u_{z}}{\partial z}\right] _{z=0}  \notag \\
&=&\mu \mathcal{F}^{-1}\left[ \left(
k_{1}{}^{2}+k_{2}{}^{2}-k_{t3}{}^{2}\right) f\left( k_{1},k_{2}\right)
\right.   \notag \\
&&+\left. 2k_{1}g_{x}\left( k_{1},k_{2}\right) +2k_{2}g_{y}\left(
k_{1},k_{2}\right) \right] .  \label{FT of stress tensors}
\end{eqnarray}

By giving the boundary values of $\sigma _{xz},$ $\sigma _{yz},$ and $\sigma _{zz},$ we can find $f,$ $g_{x},$ and $g_{y}$ from the equations above. If at least one of these three stresses is nonzero, we obtain
\begin{eqnarray}
f &=&\frac{1}{\eta _{0}\left( k_{1},k_{2}\right) }\left\{ \left(
k_{1}{}^{2}+k_{2}{}^{2}-k_{t3}{}^{2}\right) \mathcal{F}\left[ \frac{\sigma
_{zz}}{\mu }\right] \right.  \\
&&+\left. 2k_{t3}k_{1}\mathcal{F}\left[ \frac{\sigma _{xz}}{\mu }\right]
+2k_{t3}k_{2}\mathcal{F}\left[ \frac{\sigma _{yz}}{\mu }\right] \right\} ,
\end{eqnarray}
\begin{eqnarray}
g_{x} &=&\frac{1}{\eta _{0}\left( k_{1},k_{2}\right) }\left\{
2k_{l3}k_{t3}k_{1}\mathcal{F}\left[ \frac{\sigma _{zz}}{\mu }\right] \right. 
\\
&&+\left[ \frac{\eta _{1}\left( k_{1},k_{2}\right) }{k_{t3}}
-2k_{t3}k_{1}{}^{2}\right] \mathcal{F}\left[ \frac{\sigma _{xz}}{\mu }\right]
\\
&&-\left. \left[ \frac{\eta _{2}\left( k_{1},k_{2}\right) }{k_{t3}}
+2k_{t3}k_{1}k_{2}\right] \mathcal{F}\left[ \frac{\sigma _{yz}}{\mu }\right]
\right\} ,
\end{eqnarray}
and
\begin{eqnarray}
g_{y} &=&\frac{1}{\eta _{0}\left( k_{1},k_{2}\right) }\left\{2k_{l3}k_{t3}k_{2}\mathcal{F}\left[ \frac{\sigma _{zz}}{\mu }\right] \right. \\
&&-\left[ \frac{\eta _{2}\left( k_{2},k_{1}\right) }{k_{t3}} +2k_{t3}k_{1}k_{2}\right] \mathcal{F}\left[ \frac{\sigma _{xz}}{\mu }\right] \\
&&\left. +\left[ \frac{\eta _{1}\left( k_{2},k_{1}\right) }{k_{t3}} -2k_{t3}k_{2}{}^{2}\right] \mathcal{F}\left[ \frac{\sigma _{yz}}{\mu }\right] \right\} .
\end{eqnarray}
Here
\begin{eqnarray}
\eta _{1}\left( k_{1},k_{2}\right) &\equiv&k_{2}{}^{4}+k_{1}{}^{2}k_{2}{}^{2}+k_{1}{}^{2}k_{t3}{}^{2} \nonumber \\
&+&2k_{t3}k_{2}{}^{2}\left( 2k_{l3}-k_{t3}\right) +k_{t3}{}^{4} \\ 
\eta _{2}\left( k_{1},k_{2}\right)  &\equiv&k_{1}k_{2}\left[
k_{1}{}^{2}+k_{2}{}^{2}+k_{t3}\left( 4k_{l3}-3k_{t3}\right) \right]  \\
\eta _{0}\left( k_{1},k_{2}\right)  &\equiv&\eta _{1}\left( k_{1},k_{2}\right) + \frac{k_{1}}{k_{2}}\eta _{2}\left( k_{1},k_{2}\right) .
\end{eqnarray}

Therefore, from Eq.~(\ref{components of displacement}), we can find the displacement vector at $z=0$ in terms of the boundary stresses,
\begin{eqnarray}
\left. u_{x}\text{ }\right| _{z=0} &=&\mathcal{F}^{-1}\left[ \frac{-i}{
k_{t3}\eta _{0}\left( k_{1},k_{2}\right) }\left\{ k_{1}k_{t3}\eta _{3}\left(
k_{1},k_{2}\right) \mathcal{F}\left[ \frac{\sigma _{zz}}{\mu }\right]
\right. \right.   \notag \\
&&\left. \left. +\eta _{1}\left( k_{1},k_{2}\right) \mathcal{F}\left[ \frac{
\sigma _{xz}}{\mu }\right] -\eta _{2}\left( k_{1},k_{2}\right) \mathcal{F}
\left[ \frac{\sigma _{yz}}{\mu }\right] \right\} \right]   \notag \\
\left. u_{y}\text{ }\right| _{z=0} &=&\mathcal{F}^{-1}\left[ \frac{-i}{
k_{t3}\eta _{0}\left( k_{1},k_{2}\right) }\left\{ k_{2}k_{t3}\eta _{3}\left(
k_{1},k_{2}\right) \mathcal{F}\left[ \frac{\sigma _{zz}}{\mu }\right]
\right. \right.   \notag \\
&&\left. \left. +\eta _{1}\left( k_{2},k_{1}\right) \mathcal{F}\left[ \frac{
\sigma _{yz}}{\mu }\right] -\eta _{2}\left( k_{2},k_{1}\right) \mathcal{F}
\left[ \frac{\sigma _{xz}}{\mu }\right] \right\} \right]   \notag \\
\left. u_{z}\text{ }\right| _{z=0} &=&\mathcal{F}^{-1}\left[ \frac{-i}{\eta
_{0}\left( k_{1},k_{2}\right) }\left\{ \frac{\omega ^{2}}{c_{t}{}^{2}}k_{l3}
\mathcal{F}\left[ \frac{\sigma _{zz}}{\mu }\right] \right. \right.   \notag
\\
&&\left. \left. -\eta _{3}\left( k_{1},k_{2}\right) \left( k_{1}\mathcal{F}
\left[ \frac{\sigma _{xz}}{\mu }\right] +k_{2}\mathcal{F}\left[ \frac{\sigma
_{yz}}{\mu }\right] \right) \right\} \right] ,  \notag \\
&&  \label{FT of displacement}
\end{eqnarray}
with
\begin{equation}
\eta _{3}\left( k_{1},k_{2}\right) \equiv k_{1}{}^{2}+k_{2}{}^{2}+k_{t3}\left(2k_{l3}-k_{t3}\right) .
\end{equation}

\section{ENERGY TRANSMISSION FROM NANOWIRE TO BULK\label{transmission probability section}}

Now we are ready to calculate the transmission probability, defined as the ratio of transmitted to incident elastic energy flux, for each of the four gapless branches. The energy current $I$ can be expressed as\cite{Blencowe}
\begin{equation}
I = \left\langle \int_{s}dxdy\left( \frac{\partial u_{x}}{\partial t}\sigma_{xz}+\frac{\partial u_{y}}{\partial t}\sigma _{yz}+\frac{\partial u_{z}}{\partial t}\sigma _{zz}\right) 
\right\rangle _{z=0}  ,
\end{equation}
where $\left\langle ...\right\rangle $ represents a time average and $\int_{s}dxdy$ is the surface integral over the $z=0$ cross-section of the wire. In conventional complex notation for waves, this becomes
\begin{equation}
I= -\frac{\omega }{2}\text{ Im}\int_{s}dxdy\left( u_{x}\sigma _{xz}^{\ast}+u_{y}\sigma _{yz}^{\ast }+u_{z}\sigma _{zz}^{\ast }\right) _{z=0}.
\label{energy current in xyz}
\end{equation}
We will calculate the energy current for the different branches separately.

\subsection{Torsional branch}

The stress distribution (\ref{torsional BC}) in rectangular coordinates is, for $\sqrt{x^{2}+y^{2}}<b$,
\begin{eqnarray}
&&
\begin{array}{c}
\sigma _{zz}=0
\end{array} \notag \\
&&
\begin{array}{l}
\sigma _{xz}=-2i\mu kye^{-i\omega t} \\ 
\sigma _{yz}=2i\mu kxe^{-i\omega t},
\end{array}
\label{torsional BC in xy} 
\end{eqnarray}
and zero for $\sqrt{x^{2}+y^{2}} \ge b$. Using Eq.~(\ref{FT of displacement}) we obtain
\begin{eqnarray}
I &=&-\frac{\omega }{2}\text{ Im}\int_{s}dxdy\left( u_{x}\sigma _{xz}^{\ast}+u_{y}\sigma _{yz}^{\ast }\right) _{z=0} \\
&=&\frac{\omega }{4\pi }\text{ Re}\int \frac{dk_{1}dk_{2}}{k_{t3}\eta_{0}\left( k_{1},k_{2}\right) }\int_{s}dxdye^{-i\left( k_{1}x+k_{2}y\right) } \nonumber \\
&\times&\left\{ \sigma _{xz}^{\ast }\left( \eta _{1}\left( k_{1},k_{2}\right) \mathcal{F}\left[ \frac{\sigma _{xz}}{\mu }\right] -\eta _{2}\left(
k_{1},k_{2}\right) \mathcal{F}\left[ \frac{\sigma _{yz}}{\mu }\right] \right) \right. \nonumber \\
&+&\left. \sigma _{yz}^{\ast }\left( \eta _{1}\left( k_{2},k_{1}\right) \mathcal{F}\left[ \frac{\sigma _{yz}}{\mu }\right] -\eta _{2}\left(
k_{2},k_{1}\right) \mathcal{F}\left[ \frac{\sigma _{xz}}{\mu }\right] \right) \right\} \nonumber \\
&=& 2\mu k^{2}\omega \text{ Re}\int \frac{dk_{1}dk_{2}}{k_{t3}\eta _{0}\left( k_{1},k_{2}\right) } \nonumber \\
&\times&\left\{ \mathcal{F}\left[ y\right] ^{\ast }\left( \eta _{1}\left(k_{1},k_{2}\right) \mathcal{F}\left[ y\right] +\eta _{2}\left(k_{1},k_{2}\right) \mathcal{F}\left[ x\right] \right) 
\right. \nonumber \\
&+&\left.\mathcal{F}\left[ x\right] ^{\ast }\left( \eta _{1}\left(
k_{2},k_{1}\right) \mathcal{F}\left[ x\right] +\eta _{2}\left(
k_{2},k_{1}\right) \mathcal{F}\left[ y\right] \right) \right\} .
\end{eqnarray}
Expanding the Fourier transforms $\mathcal{F}\left[x\right] $\ and $\mathcal{F}\left[ y\right] $ for small $kb$, and keeping only the leading terms,
\begin{eqnarray}
\frac{1}{2\pi }\int_{s}xe^{-i\left( k_{1}x+k_{2}y\right) }dxdy &\simeq &
\frac{-ik_{1}b^{4}}{8} \\
\frac{1}{2\pi }\int_{s}ye^{-i\left( k_{1}x+k_{2}y\right) }dxdy &\simeq &
\frac{-ik_{2}b^{4}}{8},
\end{eqnarray}
we have
\begin{eqnarray}
I &=&\frac{\mu b^{8}k^{2}\omega }{32}\text{ Re}\int \frac{dk_{1}dk_{2}}{
k_{t3}\eta _{0}\left( k_{1},k_{2}\right) } \nonumber \\
&\times&\left\{ k_{2}{}^{2}\left[ \eta _{1}\left( k_{1},k_{2}\right) +\frac{k_{1}}{k_{2}}\eta _{2}\left( k_{1},k_{2}\right) \right] \right. \nonumber \\
&&+\left. k_{1}{}^{2}\left[ \eta _{1}\left( k_{2},k_{1}\right) +\frac{k_{2}}{
k_{1}}\eta _{2}\left( k_{2},k_{1}\right) \right] \right\} \\
&=&\frac{\mu b^{8}k^{2}\omega }{32}\text{ Re}\int dk_{1}dk_{2}\frac{
k_{2}{}^{2}+k_{1}{}^{2}}{k_{t3}} \\
&=&\frac{\pi }{24}\mu b^{8}k^{5}\omega .
\end{eqnarray}

By normalizing $I$ with the energy current $\frac{\pi }{4}\mu b^{4}k\omega $ of the incident torsional wave, we obtain the transmission probability
\begin{equation}
\mathcal{T} =\frac{1}{6}b^{4}k^{4}.
\end{equation}

\subsection{Longitudinal branch}

Using Eq.~(\ref{longitudinal BC}),
\begin{equation}
I=-\frac{\omega }{2}\text{ Im}\int_{0}^{b}rdr\int_{0}^{2\pi }d\theta \left(u_{z}\sigma _{zz}^{\ast }\right) _{z=0}.
\end{equation}
Then to leading order in $kb$,
\begin{eqnarray}
I &=&\frac{\omega }{8\pi ^{2}\mu }\left| \int_{0}^{b} \! rdr \! \int_{0}^{2\pi} \! d\theta \sigma _{zz}\right| ^{2} \! \text{Re} \! \int \! dk_{1}dk_{2}\frac{\omega^{2}k_{l3}}{c_{t}{}^{2}
\eta _{0}\left( k_{1},k_{2}\right) } \nonumber  \\
&=&\frac{\mu }{2c_{t}}\left( \frac{c_{0}{}^{2}}{c_{t}{}^{2}}b^{2}k\omega\right) ^{2}\text{Re}\int dk_{1}dk_{2}\frac{\omega k_{l3}}{c_{t}\eta_{0}\left( k_{1},k_{2}\right) }.
\end{eqnarray}
Normalizing by the power $\frac{\pi }{2}\mu \frac{c_{0}{}^{2}}{c_{t}{}^{2}}b^{2}k\omega$ carried by the incident wave leads to
\begin{eqnarray}
\mathcal{T}\left( k\right) &=&\frac{c_{0}{}^{2}}{\pi c_{t}{}^{3}}b^{2}k\omega \text{ Re}\int dk_{1}dk_{2}\frac{\omega k_{l3}}{c_{t}\eta
_{0}\left( k_{1},k_{2}\right) }  \\
&=&t_{l}b^{2}k^{2},
\end{eqnarray}
where
\begin{equation}
t_{l} \equiv \frac{c_{0}{}^{3}}{\pi c_{t}{}^{3}}\text{ Re}\int dk_{1}dk_{2}\frac{\omega k_{l3}}{c_{t}\eta _{0}\left( k_{1},k_{2}\right) } .
\label{tl}
\end{equation}
Assuming the materials to be made of Si, we have $p=0.694$ and
\begin{eqnarray}
\mathcal{T} &=&1.91b^{2}k^{2}   \\
&=&0.923\left( \frac{b\omega }{c_{t}}\right) ^{2} \! .
\end{eqnarray}

\subsection{Flexural branches}

By using flexural stress distribution (\ref{flexural BC}), we find that to leading order in $kb$ the energy current is
\begin{eqnarray}
I &=&\frac{\omega }{2}\text{ Re}\int_{s}dxdy\mathcal{F}^{-1}\left\{ \frac{\eta _{1}\left( k_{1},k_{2}\right) }{k_{t3}\eta _{0}\left(
k_{1},k_{2}\right) }\mathcal{F}\left[ \frac{\sigma _{xz}}{\mu }\right] \right\} \sigma _{xz}^{\ast } \nonumber \\
&=&\frac{\omega }{8\pi ^{2}\mu }\left| \int_{s}\sigma _{xz}dxdy\right| ^{2} \text{Re}\int dk_{1}dk_{2}\frac{\eta _{1}\left( k_{1},k_{2}\right) }{
k_{t3}\eta _{0}\left( k_{1},k_{2}\right) } \nonumber \\
&=&\frac{\mu \omega k^{6}}{32\pi ^{2}}\left[ \int_{s}dxdy\left( 1+\frac{c_{0}{}^{2}}{c_{t}{}^{2}}\right) \left( b^{2}-x^{2}\right) \right. \nonumber  \\
&+&\left. \left( \frac{c_{0}{}^{2}}{c_{t}{}^{2}}-3\right) y^{2}\right] ^{2}\text{Re}\int dk_{1}dk_{2}\frac{\eta _{1}\left( k_{1},k_{2}\right) }{
k_{t3}\eta _{0}\left( k_{1},k_{2}\right) } \nonumber \\
&=&\frac{\mu }{32}\left( \frac{c_{0}{}^{2}}{c_{t}{}^{2}}b^{4}k^{3}\right)^{2}\omega \text{ Re}\int dk_{1}dk_{2}\frac{\eta _{1}\left(
k_{1},k_{2}\right) }{k_{t3}\eta _{0}\left( k_{1},k_{2}\right) }.
\end{eqnarray}
Normalizing by the energy current of the incident wave 
\begin{eqnarray}
I_{\rm in} &=&-\frac{\omega }{2}\text{ Im}\int_{s}dxdy\left( \sigma _{xz}^{\ast}-ikx\sigma _{zz}^{\ast }\right) _{z=0} \\
&=&\frac{\pi \mu }{4}\frac{c_{0}{}^{2}}{c_{t}{}^{2}}b^{4}k^{3}\omega,
\end{eqnarray}
where we have used Eqs.~(\ref{flexural displacement}) and (\ref{flexural stress tensors}), the transmission probability is found to be
\begin{eqnarray}
\mathcal{T} &=&\frac{1}{8\pi }\frac{c_{0}{}^{2}}{c_{t}{}^{3}}b^{4}k^{3}\omega \text{ Re}\int dk_{1}dk_{2}\frac{c_{t}\eta _{1}\left(k_{1},k_{2}\right) }{\omega k_{t3}\eta _{0}
\left( k_{1},k_{2}\right) } \\
&=&t_{f}b^{5}k^{5}
\end{eqnarray}
where
\begin{equation}
t_{f}\equiv \frac{c_{0}{}^{3}}{16\pi c_{t}{}^{3}}\text{ Re}\int dk_{1}dk_{2}\frac{
c_{t}\eta _{1}\left( k_{1},k_{2}\right) }{\omega k_{t3}\eta _{0}\left(
k_{1},k_{2}\right) }.  \label{tf}
\end{equation}

Using $p=0.694$ for Si, the transmission probability becomes
\begin{eqnarray}
\mathcal{T} &=&0.268 \, b^{5}k^{5}  \\
&=& 0.609\left( \frac{b\omega }{c_{t}}\right) ^{5/2}
\end{eqnarray}
Because of the cylindrical symmetry, the $y$-polarized flexural branch has the same transmission probability as the $x$-polarized branch.

\section{CONCLUSIONS\label{conclusion section}}

On the left side of Table \ref{summary table} we summarize the transmission probability results calculated above, as well as the low-frequency dispersion relations of the four gapless modes. For comparison with the results of Cross and Lifshitz\cite{Cross and Lifshitz} for a rectangular wire connected to a thin plate, we reproduce their results on the right side of this Table. In each case there are four gapless acoustic modes: one torsional, one longitudinal (or compressional) and two flexural bending modes. Also, the form of the dispersion relations are the same for both wires. For all branches the transmission probability to a three-dimensional bulk solid has a higher-order frequency dependence. This is at least partially a consequence of the higher vibrational density of states in the three-dimensional system as compared to a plate: For the longitudinal and $x$-polarized flexural branches, $\mathcal{T}$ is one order higher in $\omega$, consistent with the density of states enhancement.\cite{DOS footnote}

The phonon transmission probabilities can be used to calculate the mesoscopic thermal conductance between an equilibrated wire and bulk. According to the thermal Landauer formula,\cite{Rego and Kirczenow,Blencowe,Angelescu} a {\it total} transmission probability ${\sf T}(\omega)$ varying at low frequency as $\omega^\gamma$ will lead to a low-temperature thermal conductance varying with temperature as $G_{\rm th} \propto T^{\gamma+1}.$ In our case, ${\sf T}(\omega)$ is a sum of the ${\cal T}$ for each channel. The thermal conductance between an equilibrated cylindrical wire nonadiabatically coupled to a bulk solid should therefore vanish with temperature as $T^3 \! .$

Finally we comment on the applicability of our results to nanoscale phonon experiments, which do not consider infinitely long wires and perfectly sharp corners. For our theory to be valid, the wire must be longer than the sound wavelength, and the characteristic radii of curvature at the junction must be much smaller than this wavelength. Therefore, because of the first condition, our results will become invalid in the extreme low-temperature limit, and the conductance will crossover from our predicted $T^3$ scaling to some other behavior.

 \begin{widetext}

\begin{table}[h]
\caption{(left) Dispersion relations $\protect\omega \left( k\right) $ of the low-frequency vibrational modes in a cylindrical nanowire, and transmission probabilities ${\cal T}$ through the junction with a three-dimensional bulk insulator, as a function of both $k$ and $\omega$. $t_{l}$ and$\ t_{f}$ are constants defined in Eqs.~(\ref{tl}) and (\ref{tf}). In the low-frequency limit the total transmission probability vanishes as $\omega^2 \! ,$ the transport being dominated by the longitudinal channel. (right) Same quantities for a rectangular wire connected to a thin plate, reproduced from Ref.~[\onlinecite{Cross and Lifshitz}]. Here $I_{1}$ and $I_{2}$ are Poisson-ratio-dependent numbers.}
\begin{center}
\par
\begin{tabular}{|c|c|c|c|c|c|c|c|}
\hline
\multicolumn{4}{|c}{cylindrical nanowire (radius $b$)} & 
\multicolumn{4}{|c|}{rectangular nanowire (width $b$, thickness $d$)} \\ 
\multicolumn{4}{|c}{$\longrightarrow $ semi-infinite space (3D solid)} & \multicolumn{4}{|c|}{$\longrightarrow $ thin plate
(2D plate of thickness $d$)} \\ \hline
branch & $\omega \left( k\right) $ & $\mathcal{T}\left( k\right) $ & $\mathcal{T}\left( \omega \right) $ & branch & $\omega \left( k\right) $ & $\mathcal{T}
\left( k\right) $ & $\mathcal{T}\left( \omega \right) $ \\ \hline
\multicolumn{1}{|l|}{torsional} & $c_{t}k$ & $\frac{1}{6}\left( bk\right)
^{4}$ & $\frac{1}{6}\left( \frac{b\omega }{c_{t}}\right) ^{4}$ & 
\multicolumn{1}{|l|}{torsional} & $\frac{2d}{b}c_{t}k$ & $I_{2}bk$ & $I_{2}
\frac{b}{2d}\frac{b\omega }{c_{t}}$ \\ \hline
\multicolumn{1}{|l|}{longitudinal} & $c_{0}k$ & $t_{l}\left( bk\right) ^{2}$
& $t_{l}\left( \frac{b\omega }{c_{0}}\right) ^{2}$ & \multicolumn{1}{|l|}{
compressional} & $c_{0}k$ & $4bk$ & $4\frac{b\omega }{c_{0}}$ \\ \hline
\multicolumn{1}{|l|}{flexural ($x$-direction bending)} & $\frac{1}{2}
c_{0}bk^{2}$ & $t_{f}\left( bk\right) ^{5}$ & $4\sqrt{2}t_{f}\left( \frac{
b\omega }{c_{0}}\right) ^{5/2}$ & \multicolumn{1}{|l|}{in-plane bending} & $
\frac{\sqrt{3}}{6}c_{0}bk^{2}$ & $\frac{1}{3}\left( bk\right) ^{3}$ & $2\sqrt
[4]{\frac{4}{3}}\left( \frac{b\omega }{c_{0}}\right) ^{3/2}$ \\ \hline
\multicolumn{1}{|l|}{flexural ($y$-direction bending)} & \multicolumn{3}{|c}{
same as $x$ direction} & \multicolumn{1}{|l|}{normal-plane bending} & $
\frac{\sqrt{3}}{6}c_{0}dk^{2}$ & $I_{1}bk$ & $I_{1}\sqrt[4]{12}\left( \frac{
b^{2}\omega }{dc_{0}}\right) ^{1/2}$ \\ \hline
\end{tabular}
\end{center}
\label{summary table}
\end{table}

\acknowledgements

This work was supported by the National Science Foundation under CAREER Grant No.~DMR-0093217. Acknowledgment is also made to the Donors of the American Chemical Society Petroleum Research Fund, for partial support of this research.

\appendix

\section{MOMENTUM CONSERVATION AND HARD-WALL BOUNDARY CONDITION}

Here we use linear and angular momentum conservation to derive Eq.~(\ref{hard-wall boundary conditions}) in the $B \gg b$ limit. First we equate the torques produced by the thin and thick wires on each other. From the rotational stress
\begin{eqnarray}
\sigma _{\theta z} &=&\mu \left( \frac{\partial u_{\theta }}{\partial z}+ \frac{1}{r}\frac{\partial u_{z}}{\partial \theta }\right) \\
&=&\left\{ 
\begin{array}{l}
\sigma _{i\theta z}(r,\theta) e^{i\left( kz-\omega t\right) }+R_{ij}\sigma _{j\theta z}^{\ast }(r,\theta) e^{-i\left( kz+\omega t\right) }  \ \text{for} \ z<0, \\  
T_{ij}\sigma _{j\theta z}(r,\theta) e^{i\left( kz-\omega t\right) } \ \text{for} \ z>0,
\end{array} \right. 
\end{eqnarray}
the torsional torque $\tau =\int_{s}rdrd\theta r\sigma _{\theta z}$ is found to be
\begin{equation}
\tau =\left\{ 
\begin{array}{cc}
\frac{i\pi }{2}b^{4}\left( \delta _{i1}e^{ikz}-R_{i1}e^{-ikz}\right) \mu ke^{-i\omega t} & \text{for} \ z=0^- \\ &  \\ 
\frac{i\pi }{2}B^{4}T_{i1}\mu ke^{i\left( kz-\omega t\right) } & \text{for} \ z=0^+.
\end{array}
\right. 
\end{equation}
Only torsional mode contributes to the torque. 

By equating these torques we have 
\begin{equation}
b^{4}\left( \delta _{i1}-R_{i1}\right) =B^{4}T_{i1}.
\label{conservation of torque}
\end{equation}
Combining this result with Eq.~(\ref{contiuum of displacement vector}) leads to
\begin{equation}
R_{i1}=T_{i1}=0,\text{ \ \ \ \ \ }\ i\neq 1
\end{equation}
\begin{equation}
R_{11}=-\frac{B^{4}-b^{4}}{B^{4}+b^{4}}\text{ },\text{ \ \ \ \ }T_{11}=\frac{2b^{4}}{B^{4}+b^{4}}\text{ }.
\end{equation}
Taking the $B/b \rightarrow \infty$ limit then leads to $R_{11}\rightarrow -1$ and $T_{11}\rightarrow 0$. 

An analogous result for $R_{i2}$\ and $T_{i2}$ can be derived by considering the total force in $z$ direction,
\begin{eqnarray}
F_{z} &=&\int rdrd\theta \sigma _{zz} \\
&=&\left\{ \begin{array}{cc}
i\pi b^{2}\left( \delta _{i2}e^{ikz}-R_{i2}e^{-ikz}\right) \frac{c_{0}{}^{2}}{c_{t}{}^{2}}\mu ke^{-i\omega t} & \text{for} \ z<0, \\ 
&  \\ i\pi B^{2}T_{ij}\frac{c_{0}{}^{2}}{c_{t}{}^{2}}\mu ke^{i\left( kz-\omega t\right) } & \text{for} \ z>0,
\end{array} \right. 
\end{eqnarray}
which gives
\begin{equation}
b^{2}\left( \delta _{i2}-R_{i2}\right) =B^{2}T_{i2},
\label{conservation of fz}
\end{equation}
since only longitudinal mode has a nonzero $F_{z}$ . Then
\begin{equation}
R_{i2}=T_{i2}=0,\text{ \ }\ i\neq 2
\end{equation}
\begin{eqnarray}
R_{22} &=&-\frac{B^{2}-b^{2}}{B^{2}+b^{2}}\rightarrow -1 \\
T_{22} &=&\frac{2b^{2}}{B^{2}+b^{2}}\text{ \ }\rightarrow \text{ }0,
\end{eqnarray}
for $B/b \rightarrow \infty$. 

By further considering the conservation of momentum in the $x$ and $y$ directions, it is not difficult to derive the result quoted in Eq.~(\ref{hard-wall boundary conditions}).

\end{widetext}


\begin{thebibliography}{99}
\bibitem{adiabatic footnote} Adiabaticity was only marginally satisfied at the lowest temperatures reached in the experiment of Ref.~[\onlinecite{Yung etal quantization}].
\bibitem{DOS footnote} For the other two branches, the motion of the displacement field in the thin plate is normal to the surface of the plate, and it is not clear whether one can make such a comparison.
\end{thebibliography}
\end{document}